\documentclass[prd,aps,showpacs,nofootinbib,eqsecnum,onecolumn,groupedaddress,
amssymb]{revtex4}
\usepackage{graphicx}
\usepackage[english]{babel}
\usepackage{amsmath}
\usepackage{amssymb}
\usepackage{amsfonts}
\usepackage{latexsym}
\usepackage{color,amsxtra}
\usepackage{epsf}
\usepackage{enumerate}
\usepackage{hhline}
\usepackage{array}
\usepackage{tabularx}
\usepackage{subfigure}
\usepackage{fancyhdr}
\usepackage{mathrsfs}
\newcommand{\be}{\begin{equation}}
\newcommand{\ee}{\end{equation}}
\newcommand{\bea}{\begin{eqnarray}}
\newcommand{\eea}{\end{eqnarray}}
\newcommand{\beaa}{\begin{eqnarray*}}
\newcommand{\eeaa}{\end{eqnarray*}}

\newcommand{\e}{\mathrm{e}}

\newcommand{\Eqn}[1]{&\hspace{-0.2em}#1\hspace{-0.2em}&}
\def\Vec#1{\mbox{\boldmath $#1$}}

\def\be{\begin{equation}}
\def\ee{\end{equation}}
\def\bea{\begin{eqnarray}}
\def\eea{\end{eqnarray}}

\def\e{\mathrm{e}}

\begin{document}

\title{Inflation in a viscous fluid model}

\author{Kazuharu Bamba$^{1}$ and Sergei D. Odintsov$^{2, 3}$
}
\affiliation{
$^1$Division of Human Support System, Faculty of Symbiotic Systems Science, Fukushima University, Fukushima 960-1296, Japan\\
$^2$Institut de Ciencies de lEspai (IEEC-CSIC), 
Campus UAB, Carrer de Can Magrans, s/n 
08193 Cerdanyola del Valles, Barcelona, Spain\\
$^3$Instituci\'{o} Catalana de Recerca i Estudis Avan\c{c}ats
(ICREA), Passeig Llu\'{i}s Companys, 23 08010 Barcelona, Spain
}


\begin{abstract} 
We explore a fluid description of the inflationary universe. 
In particular, we investigate a fluid model in which the equation of state (EoS) for a fluid includes bulk viscosity. 
We find that the three observables of inflationary cosmology: 
the spectral index of the curvature perturbations, 
the tensor-to-scalar ratio of the density perturbations, 
and the running of the spectral index, 
can be consistent with the recent Planck results. 
We also reconstruct the explicit EoS for a fluid 
from the spectral index of the curvature perturbations 
compatible with the Planck analysis. 
In the reconstructed models of a fluid, 
the tensor-to-scalar ratio of the density perturbations 
can satisfy the constraints obtained from the Planck satellite. 
The running of the spectral index can explain the Planck data. 
In addition, it is demonstrated that 
in the reconstructed models of a fluid, 
the graceful exit from inflation can be realized. 
Moreover, we show that the singular inflation can occur in a fluid model. 
Furthermore, we show that a fluid description of inflation can be equivalent to the description of inflation in terms of scalar field theories. 
\end{abstract}

\pacs{98.80.-k, 98.80.Cq, 04.50.Kd, 12.60.-i}
\hspace{15.0cm} FU-PCG-07

\maketitle

\def\thesection{\Roman{section}}
\def\theequation{\Roman{section}.\arabic{equation}}

\section{Introduction}

The various precise properties on inflation in the early universe~\cite{Inflation, N-I, Starobinsky:1980te} has been revealed by the recent cosmological observations on the anisotropy of the cosmic microwave background (CMB) radiation 
such as the Planck satellite~\cite{Planck:2015xua, Ade:2015lrj} and 
the BICEP2 experiment~\cite{Ade:2014xna, Ade:2015tva}, 
in addition to the Wilkinson Microwave anisotropy probe (WMAP)~\cite{Komatsu:2010fb, Hinshaw:2012aka}. 
The nature of inflation can be known from the spectrum of the primordial density perturbations~\cite{IM-PDP}. 

Recently, in Refs.~\cite{Bamba:2014daa, Bamba:2014wda}, 
the convenient fluid description of the inflationary universe has been proposed in terms of scalar field theories, fluid models~\cite{Bamba:2012cp}, and $F(R)$ gravity theories~\cite{R-NO-CF-CD}. 
Especially, the three observables of inflationary models, 
i.e., the spectral index of the curvature perturbations, 
the tensor-to-scalar ratio of the density perturbations, 
and the running of the spectral index, 
have been represented by using quantities in scalar field theories, 
fluid models, and $F(R)$ gravity theories. 
Fluid models have been applied to cosmological issues 
such as inflation~\cite{Barrow:1994nx, Brevik:2011mm, Wang:2013uka, 
Elizalde:2014ova, Brevik:2014eya, Myrzakulov:2014kta, Myrzakul:2015cua} 
and dark energy~\cite{Nojiri:2005sr} 
(for recent reviews, see~\cite{Bamba:2012cp, Brevik:2014cxa}). 

In this paper, by extending the preceding investigations of 
a fluid description~\cite{Bamba:2014wda} for the inflationary universe, 
we derive the slow-roll conditions and construct the formulae of 
the observables for inflationary models of 
the spectral index $n_\mathrm{s}$ of the curvature perturbations, 
the tensor-to-scalar ratio $r$ of the density perturbations, and 
the running $\alpha_\mathrm{s}$ of the spectral index. 
We analyze the equation of state (EoS) for a fluid including bulk viscosity 
and examine a fluid model in which $n_\mathrm{s}$, $r$, 
and $\alpha_\mathrm{s}$ 
can be compatible with the recent Planck results. 

In addition, we explicitly reconstruct the EoS for a fluid 
from the spectral index $n_\mathrm{s}$ of the curvature perturbations. 
Particularly, we use 
the expression of $n_\mathrm{s}$ as a function of the number of 
$e$-folds $N$ during inflation 
in the inflationary models including 
the Starobinsky inflation~\cite{Starobinsky:1980te}, 
from which the value of $n_\mathrm{s}$ consistent the recent Planck analysis 
can be obtained. 
This reconstruction method for scalar field theories has been 
proposed in Ref.~\cite{Chiba:2015zpa}. 
In this work, we present the reconstruction procedure in a fluid description. 
We also show that in the reconstructed models of a fluid, 
the slow-roll inflation, i.e., the de Sitter inflation, can occur. 
In these fluid models, 
the tensor-to-scalar ratio $r$ of the density perturbations 
can meet the constraints acquired by the Planck satellite. 
The running $\alpha_\mathrm{s}$ of the spectral index can 
explain the Planck results. 
Moreover, we certify that in the reconstructed models of a fluid, 
the graceful exit from inflation can be realized. 
Furthermore, we demonstrate that in the fluid models, 
the singular inflation~\cite{Barrow:2015ora, Nojiri:2015fia, Nojiri:2015wsa, Odintsov:2015jca} can occur. 
In addition, we explain the equivalence between a fluid description of inflation and the description of inflation in terms of scalar field theories. 
We use units of $k_\mathrm{B} = c = \hbar = 1$ and express the
gravitational constant $8 \pi G_\mathrm{N}$ by
${\kappa}^2 \equiv 8\pi/{M_{\mathrm{Pl}}}^2$ 
with the Planck mass of $M_{\mathrm{Pl}} = G_\mathrm{N}^{-1/2} = 1.2 \times 
10^{19}$\,\,GeV. 

The organization of the paper is the following. 
In Sec.~II, we explain a fluid description of the inflationary universe. 
In particular, we present a fluid model in which the EoS for a fluid includes bulk viscosity and show that the fluid model can explain the recent Planck results of the three observables for inflationary models. 
In Sec.~III, we reconstruct the EoS for a fluid from 
the spectral index of the curvature perturbations. 
We certify that inflation can happen in the reconstructed models of a fluid, and that the tensor-to-scalar ratio of the density perturbations can be consistent with the Planck analysis. 
In Sec.~IV, we investigate that the graceful exit from inflation 
can be realized in the fluid models reconstructed above. 
In Sec.~V, we consider the singular inflation in a fluid model. 
In Sec.~VI, we show that a fluid description of inflation can be equivalent to the description of inflation in terms of scalar field theories. 
Conclusions are presented in Sec.~VII. 
In Appendix A, the slow-roll parameters in a fluid description are given. 

\section{Fluid description of inflation}

We consider the case that the so-called slow-roll inflation driven by 
the potential $V(\phi)$ of a scalar filed $\phi$ occurs, 
which plays a roll of the inflaton field. 
We explain the procedure~\cite{Bamba:2014daa} 
to represent the slow-roll parameters 
in terms of the Hubble parameter and its derivatives of 
the number of $e$-folds during inflation. 
Furthermore, with these representations of the slow-roll parameters, 
we describe the observables of inflationary models, namely, 
the spectral index of the curvature perturbations, 
the tensor-to-scalar ratio of the density perturbations, 
and the running of the spectral index in fluid models~\cite{Bamba:2014wda}. 

\subsection{Slow-roll parameters} 

The action of $\phi$ with the Einstein-Hilbert term is given by 
\begin{equation}
S = \int d^4 x \sqrt{-g} \left( \frac{R}{2\kappa^2} 
 - \frac{1}{2}\partial_\mu \phi \partial^\mu \phi - V(\phi) \right)\, . 
\label{eq:2.1}
\end{equation}
Here, $g$ is the determinant of the metric $g_{\mu\nu}$ and 
$R$ is the scalar curvature.  
For the slow-roll inflation, 
the spectral index $n_\mathrm{s}$ of the curvature perturbations 
(i.e., the scalar mode of the density perturbations), 
the tensor-to-scalar ratio $r$ of the density perturbations, 
and the running of the spectral index 
$\alpha_\mathrm{s} \equiv d n_\mathrm{s}/d \ln k$, 
where $k$ is the absolute value of the wave number $\Vec{k}$, 
are written as
\begin{equation}
n_\mathrm{s} - 1 = - 6 \epsilon + 2 \eta\, , 
\quad 
r = 16 \epsilon \, , 
\quad 
\alpha_\mathrm{s} = 16\epsilon \eta - 24 \epsilon^2 - 2 \xi^2\, ,
\label{eq:2.3}
\end{equation}
where  $\epsilon$, $\eta$, and $\xi$ are 
the slow-roll parameters, defined as 
\begin{equation}
\epsilon \equiv 
\frac{1}{2\kappa^2} \left( \frac{V'(\phi)}{V(\phi)} \right)^2\, , 
\quad 
\eta \equiv \frac{1}{\kappa^2} \frac{V''(\phi)}{V(\phi)}\, , 
\quad 
\xi^2 \equiv \frac{1}{\kappa^4} \frac{V'(\phi) V'''(\phi)}{\left(V(\phi)\right)^2}\, .
\label{eq:2.2}
\end{equation}
Here, the prime shows the derivative with respect to $\phi$ as 
$V' (\phi) \equiv d V(\phi)/d \phi$. 
Throughout this paper, the prime denotes the derivative with respect to 
the argument of the function, to which the prime operates. 

We take the flat Friedmann-Lema\^{i}tre-Robertson-Walker (FLRW) metric 
$ds^2 = -dt^2 + a^2(t) \sum_{i=1,2,3}\left(dx^i\right)^2$, 
where $a(t)$ is the scale factor. 
The Hubble parameter is 
defined by $H \equiv \dot{a}/a$, where the dot 
means the time derivative. 

We express the slow-roll parameters in terms of $H$, 
which can be represented as $H=H(N)$, namely, as 
a function of the number of $e$-folds $N$ 
during inflation, defined as 
$N \equiv \ln \left(a_\mathrm{f}/a_\mathrm{i}\right) 
= \int_{t_\mathrm{i}}^{t_\mathrm{f}} H dt$, 
where $a_\mathrm{i}$ and $a_\mathrm{f}$ 
are the values of the scale factor $a$ at the initial time $t_\mathrm{i}$ and 
the end time $t_\mathrm{f}$ of inflation, respectively. 
To execute this task, with a new scalar field $\varphi$, 
we redefine $\phi$ as $\phi = \phi (\varphi)$, 
where $\varphi$ is identified with $N$. 
We introduce a positive quantity $\omega(\varphi) \, (>0)$ defined as 
$\omega(\varphi) \equiv \left( d\phi/d\varphi \right)^2$, 
and represent $V$ as a function of $\varphi$, i.e., 
$V(\varphi)\equiv V\left(\phi\left(\varphi\right)\right)$. 
In the FLRW background, we derive 
the gravitational equations and rewrite them by using 
$\omega(\varphi)$ and $V(\varphi)$. 
By solving the gravitational equations with respect to 
$\omega(\varphi)$ and $V(\varphi)$, we obtain~\cite{Bamba:2014daa}
\begin{equation}
\omega(\varphi) 
= - \left. \frac{2}{\kappa^2}\frac{H'(N)}{H(N)}\right|_{N=\varphi}\, , 
\quad 
V(\varphi) = \left. 
\frac{1}{\kappa^2} \left(H(N)\right)^2 \left( 3 + \frac{H'(N)}{H(N)} \right)
\right|_{N=\varphi}\, ,
\label{eq:2.4}
\end{equation}
with $H'(N) \equiv dH(N)/dN$. 
Here, the representations of $H=H(N)$ and $\varphi = N$ 
are acquired as solutions for the equation of motion of 
$\phi$ or $\varphi$, and the gravitational field equations. 
It is seen from the first equation in (\ref{eq:2.4}) 
that since $\omega(\varphi)>0$, we have $H'(N)<0$. 
The slow-roll parameters in (\ref{eq:2.2}) 
can be rewritten by using 
$\omega(\varphi)$ and $V(\varphi)$. 
Accordingly, through the expressions of $\omega(\varphi)$ and $V(\varphi)$ 
in (\ref{eq:2.4}), 
the slow-roll parameters can be described in terms of $H(N)$ 
and its derivatives with respect to $N$. 
The resultant expressions have been given in Ref.~\cite{Bamba:2014daa}.

\subsection{Representation of a fluid} 

For a general fluid model, the EoS is given by 
\begin{equation}
P(N) = - \rho(N) + f(\rho) \,, 
\label{eq:2.5}
\end{equation}
where $\rho$ is the energy density of a fluid, 
$P$ is its pressure, and 
$f(\rho)$ is an arbitrary function of $\rho$. 
In the flat FLRW background, for such a fluid model, 
the gravitational field equations read 
\begin{eqnarray}
\frac{3}{\kappa^2} \left(H (N)\right)^2 \Eqn{=} \rho \,, 
\label{eq:FR16-3-IIB1} \\
- \frac{2}{\kappa^2} H(N) H'(N) \Eqn{=} \rho + P \,,
\label{eq:FR16-3-IIB2}
\end{eqnarray}
Since the EoS can be expressed as $\rho(N) + P(N) = f(\rho)$, 
the second gravitational equation is rewritten to 
$-\left(2/\kappa^2 \right) H(N) H'(N) = f(\rho)$. 
Similarly, with the expression of the EoS shown above, 
the conservation law $0=\rho'(N) + 3 \left(\rho(N) + P(N)\right)$ 
becomes $0 = \rho'(N) + 3 f(\rho)$, 
where $\rho'(N) \equiv d\rho(N)/dN$. {}From these second gravitational 
equation and conservation law, we acquire 
\begin{equation}
\frac{2}{\kappa^2} \left(H (N)\right)^2 
\left[ \left( \frac{H'(N)}{H(N)}\right)^2 + \frac{H''(N)}{H(N)} \right] 
= 3 f'(\rho) f(\rho) \,,
\label{eq:II-6}
\end{equation}
with $f'(\rho) \equiv df(\rho)/d\rho$. 
Owing to this equation, it is possible to express 
$H(N)$ and its derivatives with respect to $N$ only with 
$\rho(N)$ and $f(\rho(N))$. 
Therefore, the slow-roll parameters can be described 
in terms of $\rho(N)$ and $f(\rho(N))$, as is presented in Appendix A. 
As a result, by substituting the representations of the slow-roll parameters 
in Appendix A into the expressions of 
observables of the inflationary models in (\ref{eq:2.2}), 
we obtain the fluid description of 
$n_\mathrm{s}$, $r$, and $\alpha_\mathrm{s}$. 
In Ref.~\cite{Bamba:2014wda}, 
the explicit expressions of $n_\mathrm{s}$, $r$, and $\alpha_\mathrm{s}$ 
have been shown\footnote{The other way to describe $\alpha_\mathrm{s}$ 
has been examined in Ref.~\cite{Bassett:2005xm}.}.

The form of the EoS for a fluid can also be represented as 
$w(N) \equiv P(N)/\rho(N) = -1+ f(\rho)/\rho(N)$, 
from which we find $f(\rho)/\rho(N) = w(N) +1$. 
When $\left| f(\rho)/\rho(N) \right| \ll 1$, and 
$f(\rho)$ and $\rho$ varies very slowly in the inflationary 
stage, the approximate expressions of 
$n_\mathrm{s}$, $r$, and $\alpha_\mathrm{s}$ read~\cite{Bamba:2014wda}
\begin{eqnarray} 
(n_\mathrm{s}, r, \alpha_\mathrm{s}) \Eqn{\approx} 
(1-6\frac{f(\rho)}{\rho(N)}\,, \, 
24 \frac{f(\rho)}{\rho(N)}\,, \, -9 \left(\frac{f(\rho)}{\rho(N)}\right)^2) 
\label{eq:2.6} \\ 
\Eqn{=} (1-6\left(w(N) +1\right)\,, \, 24\left(w(N) +1\right)\,, \, -9\left(w(N) +1\right)^2)\,,
\label{eq:2.7}
\end{eqnarray}
where in deriving (\ref{eq:2.7}), we have used the relation 
$f(\rho)/\rho(N) = w(N) +1$.

\subsection{Fluid model in which the EoS for a fluid includes bulk viscosity} 

We investigate a fluid with the following EoS
\begin{equation}
P = -\rho + A \rho^{\beta}+\zeta(H) \,, 
\label{eq:2.8} 
\end{equation}
where $A$ and $\beta$ are constants, and $\zeta(H)$ is bulk viscosity. 
As a specific case, we consider that $\zeta (H)$ has the following form 
\begin{equation}
\zeta(H) = \bar{\zeta} H^{\gamma} \,,
\label{eq:2.9}
\end{equation}
where $\bar{\zeta}$ and $\gamma$ are constants. 
We note that the mass dimension of $A$ is $-4\left(\beta-1\right)$, 
whereas that of $\bar{\zeta}$ is $-\left(\gamma -4 \right)$. {}From the 
Friedmann equation (\ref{eq:FR16-3-IIB1}) 
for the expanding universe ($H>0$), we get 
$H=\left(\kappa/\sqrt{3}\right) \sqrt{\rho}$. 
Hence, $\zeta(H)$ can be written as a function of 
$\rho$, namely, $\zeta(H) = \zeta(H(\rho))$. 
Consequently, by comparing Eq.~(\ref{eq:2.5}) with Eq.~(\ref{eq:2.8}) 
and using Eq.~(\ref{eq:2.9}), we acquire
\begin{equation}
f(\rho) = A \rho^{\beta}+\zeta(H(\rho)) 
= A \rho^{\beta} + \bar{\zeta} \left(\frac{\kappa}{\sqrt{3}} \right)^{\gamma} \rho^{\gamma/2} \,.
\label{eq:2.10}
\end{equation}

Here, we state a physical reason why we have considered the case that 
$\zeta(H)$ is expressed by a power in $H$ as given in Eq.~(\ref{eq:2.9}) 
and hence $f(\rho)$ is represented by the linear combination of two kinds of a power in $\rho$ as shown in Eq.~(\ref{eq:2.10}). 
It is considered that only for such a case that $f(\rho)$ is expressed by 
a series of a power in $\rho$, through a phenomenological approach, 
it is possible to analytically study the quantitative features of the EoS for 
a fluid to realize inflation in which the three observables of inflationary models, namely, the spectral index of the curvature perturbations, 
the tensor-to-scalar ratio of the density perturbations, 
and the running of the spectral index, 
can explain the recent Planck results, as is demonstrated below. 

The Planck analysis~\cite{Planck:2015xua, Ade:2015lrj} 
has shown that 
$n_{\mathrm{s}} = 0.968 \pm 0.006\, (68\%\,\mathrm{CL})$, 
$r< 0.11\, (95\%\,\mathrm{CL})$, 
and $\alpha_\mathrm{s} = -0.003 \pm 0.007\, (68\%\,\mathrm{CL})$. 
If $f(\rho)/\rho(N) = 4.35 \times 10^{-3}$, i.e., $w = -0.996$, from 
Eq.~(\ref{eq:2.6}) or Eq.~(\ref{eq:2.7}), 
we have $(n_{\mathrm{s}}, r, \alpha_\mathrm{s}) = 
(0.974, 0.104, -1.70 \times 10^{-4})$. 
These results are consistent with the Planck data. 

We here demonstrate that it is possible to 
realize these Planck results 
by choosing appropriate values of 
the model parameters in Eq.~(\ref{eq:2.8}) with Eq.~(\ref{eq:2.9}). 
In other words, we explicitly derive the values of the 
model parameters leading to $f(\rho)/\rho = 4.35 \times 10^{-3}$. 
It follows from Eq.~(\ref{eq:2.10}) that 
\begin{eqnarray}
\frac{f(\rho)}{\rho} \Eqn{=} A \rho_\mathrm{c}^{\beta-1} 
\left(\frac{\rho}{\rho_\mathrm{c}}\right)^{\beta-1}
+\bar{\zeta} \left(\frac{\kappa}{\sqrt{3}} \right)^{\gamma} 
\rho_\mathrm{c}^{\gamma/2-1} 
\left(\frac{\rho}{\rho_\mathrm{c}}\right)^{\gamma/2-1} 
\label{eq:2.11}\\
\Eqn{=} 
A \rho_\mathrm{c}^{\beta-1} 
\left(\frac{H_\mathrm{inf}}{H_0}\right)^{2\left(\beta-1\right)}
+\bar{\zeta} \left(\frac{\kappa}{\sqrt{3}} \right)^{\gamma} 
\rho_\mathrm{c}^{\gamma/2-1} 
\left(\frac{H_\mathrm{inf}}{H_0}\right)^{\gamma-2}
\,,
\label{eq:2.12}
\end{eqnarray}
where $\rho_\mathrm{c} \equiv 3H_0^2/\kappa^2 = 8.10 \times 10^{-47} \, \mathrm{GeV}^4$ is the critical density, 
$H_0 =100h \, \mathrm{km} \, \mathrm{sec}^{-1} \, \mathrm{Mpc}^{-1} = 2.13h \times 10^{-42}$GeV with $h=0.678$~\cite{Planck:2015xua} 
is the current Hubble parameter~\cite{Kolb and Turner}, 
and $H_\mathrm{inf}$ is the Hubble parameter at the inflationary stage. 
For simplicity, we set $\gamma = 2\beta$. In this case, from Eq.~(\ref{eq:2.12}), we obtain 
\begin{eqnarray}
\frac{f(\rho)}{\rho} \Eqn{=} 
J \left(\frac{H_\mathrm{inf}}{H_0}\right)^{2\left(\beta-1\right)}\,, 
\label{eq:2.13}\\
J \Eqn{\equiv} 
\left[A+ \bar{\zeta} \left(\frac{\kappa}{\sqrt{3}} \right)^{2\beta} 
\right] \rho_\mathrm{c}^{\beta-1} \,.
\label{eq:2.14}
\end{eqnarray}
For the simplest case that $\beta = 1$, when $J = 4.35 \times 10^{-3}$, 
regardless of the scale of inflation $H_\mathrm{inf}$, 
the Planck results can be realized. 
Moreover, in the case that $\beta = 2$, for example, if $(H_\mathrm{inf}, J) = (1.0 \times 10^{10} \, \mathrm{GeV}, 9.10 \times 10^{-107})$, 
$(1.0 \times 10^{5} \, \mathrm{GeV}, 9.10 \times 10^{-97})$, 
we can explain the Planck data.

\section{Reconstruction of the EoS for a fluid from the spectral index} 

In this section, we reconstruct the EoS for a fluid from 
the spectral index of the curvature perturbations. 
Such a reconstruction has been studied 
for the case of scalar field theories in Ref.~\cite{Chiba:2015zpa}. 

\subsection{Reconstruction procedure in a fluid description} 

For the slow-roll inflation in scalar field theories, whose action 
is given by Eq.~(\ref{eq:2.1}), 
the spectral index $n_\mathrm{s}$ of the curvature perturbations, 
the tensor-to-scalar ratio $r$ of the density perturbations, and 
the running $\alpha_\mathrm{s}$ of the spectral index are 
is derived as follows~\cite{Chiba:2015zpa}: 
\begin{equation} 
n_\mathrm{s} - 1 = \frac{d}{d N} \left[ \ln \left( \frac{1}{V^2(N)} \frac{dV(N)}{dN} \right) \right] \,,
\quad 
r = \frac{8}{V(N)} \frac{dV(N)}{dN} \,, 
\quad 
\alpha_\mathrm{s} = - \frac{d^2}{d N^2} \left[ \ln \left( \frac{1}{V^2(N)} 
\frac{dV(N)}{dN} \right) \right] \,. 
\label{eq:3.1}
\end{equation}

Similarly to the case of scalar field theories, 
in a fluid model, it is possible to reconstruct the EoS for a fluid from 
the spectral index $n_\mathrm{s}$ of the curvature perturbations. 
If we have the form of $n_\mathrm{s}$ as a function of $N$, 
by using the first relation in (\ref{eq:3.1}), 
we can obtain the expression of $V(N)$. 
Thanks to the Friedmann equation (\ref{eq:FR16-3-IIB1}), 
the Hubble parameter is related to $V(N)$, and hence we get $H=H(N)$. 
In a fluid model, 
with the other gravitational field equation (\ref{eq:FR16-3-IIB2}), 
we can acquire the form of $f(\rho)$ through the EoS in Eq.~(\ref{eq:2.5}).

\subsection{Inflationary models with $n_\mathrm{s} - 1 = -2/N$}

We demonstrate the reconstruction procedure in a fluid description 
by exploring the inflationary models in which $n_\mathrm{s}$ is given by 
\begin{equation} 
n_\mathrm{s} - 1 = -\frac{2}{N} \,.
\label{eq:FR16-4-IIIB-1}
\end{equation}
It is known that in the Starobinsky inflation ($R^2$ inflation)~\cite{Starobinsky:1980te}, 
$n_\mathrm{s}$ and $r$ are 
expressed as~\cite{Hinshaw:2012aka} 
the relation (\ref{eq:FR16-4-IIIB-1}) and 
$r = 12/N^2$, respectively. 
If $N=60$, we find $n_\mathrm{s}=0.967$ and $r=3.33 \times 10^{-3}$, 
which are consistent with the Planck data~\cite{Ade:2015lrj} 
(for a recent detailed review of inflation in modified gravity theories, see, 
for instance,~\cite{Bamba:2015uma}). 
The relation (\ref{eq:FR16-4-IIIB-1}) can be satisfied also in 
the chaotic inflation~\cite{Linde:1983gd} and 
the Higgs inflation with its non-minimal gravitational coupling~\cite{Higgs-Inflation}, 
or the so-called $\alpha$-attractor~\cite{Alpha-attractors}, which 
connects the Starobinsky, quadratic chaotic, and Higgs inflations. 
By combining the relation (\ref{eq:FR16-4-IIIB-1}) 
with the first equation in (\ref{eq:3.1}), we find 
\begin{equation} 
V(N) = \frac{1}{\left( C_1 /N \right) + C_2} \,, 
\label{eq:3.2}
\end{equation}
with $C_1 (>0)$ and $C_2$ constants, the mass dimension of which is four. 
For the potential $V(N)$ in Eq.~(\ref{eq:3.2}), 
from the second relation in (\ref{eq:3.1}), 
the tensor-to-scalar ratio $r$ of the density perturbations 
is expressed as 
\begin{equation} 
r = \frac{8}{N \left[1+\left(C_2/C_1\right) N \right]} \,, 
\label{eq:FR16-5-IIIB-01}
\end{equation}
Furthermore, with the third relation in (\ref{eq:3.1}), 
the running $\alpha_\mathrm{s}$ of the spectral index is 
written as 
\begin{equation} 
\alpha_\mathrm{s} = -\frac{2}{N^2} \,. 
\label{eq:FR16-7-IIIB-001}
\end{equation}
By using this expression, for $N=60$, we acquire 
$\alpha_\mathrm{s} = -5.56 \times 10^{-4}$. 
This value is consistent with the Planck analysis. 

In a fluid model, instead of the inflaton potential $V$, we use 
the EoS in Eq.~(\ref{eq:2.5}). 
In the FLRW background, the Friedmann equation 
(\ref{eq:FR16-3-IIB1}) is written as 
$\left(3/\kappa^2 \right) \left(H (N)\right)^2 = \rho (N) \approx V (N)$, 
where the last approximate equation follows from 
the slow-roll approximation that the kinetic term is much 
smaller than the potential one as 
$\left|\left(1/2\right)\dot{\phi}^2 \right| \ll V$. {}From the relation 
$\rho \approx V$ with Eq.~(\ref{eq:3.2}), we have 
\begin{equation} 
N \approx \frac{C_1 \rho}{1-C_2 \rho}\,. 
\label{eq:3.3}
\end{equation}
Furthermore, it follows from the Friedmann equation with 
the slow-roll approximation shown above that 
the Hubble parameter is expressed as 
\begin{equation} 
H(N) \approx 
\kappa \sqrt{\frac{1}{3\left[ \left( C_1 /N \right) + C_2 \right]}} \,, 
\label{eq:3.4}
\end{equation}
where $\left( C_1 /N \right) + C_2 > 0$. {}From Eq.~(\ref{eq:FR16-3-IIB1}) and 
(\ref{eq:FR16-3-IIB2}) with the Hubble parameter in Eq.~(\ref{eq:3.4}), 
we obtain
\begin{equation} 
P = -\rho - \frac{2}{\kappa^2} H(N) H'(N) 
\approx -\rho -\frac{3C_1}{N^2 \kappa^4} H^4 \,.
\label{eq:3.5}
\end{equation}
By comparing Eq.~(\ref{eq:2.5}) with Eq.~(\ref{eq:3.5}), we acquire
\begin{equation} 
f(\rho) \approx -\frac{3C_1}{N^2 \kappa^4} H^4 
\approx -\frac{1}{3C_1} \left(1-2C_2\rho +C_2^2 \rho^2 \right) \,.
\label{eq:3.6}
\end{equation}
Here, in deriving the second approximate equality, 
we have used the Friedmann equation (\ref{eq:FR16-3-IIB1}) and 
Eq.~(\ref{eq:3.3}).

\subsection{Fluid models and inflation}

Next, we explicitly show the models of a fluid, 
in which the values of $n_\mathrm{s}$ and $r$ 
are consistent with the Planck results. 
Plugging Eqs.~(\ref{eq:2.8}) and (\ref{eq:2.10}), 
we have the form of EoS for a fluid 
\begin{equation}
P = -\rho + f(\rho) 
= -\rho + A \rho^{\beta}+ \bar{\zeta} \left(\frac{\kappa}{\sqrt{3}} \right)^{\gamma} \rho^{\gamma/2}\,. 
\label{eq:3.7}  
\end{equation}
With the results in the preceding subsection, 
we decide the models parameters: $A$, $\bar{\zeta}$, $\beta$, and $\gamma$, 
in which the relation (\ref{eq:FR16-4-IIIB-1}) can be satisfied. 

\subsubsection{Case (i): $\left| C_2\rho \right| \gg 1$}

When $\left| C_2\rho \right| \gg 1$, from Eq.~(\ref{eq:3.6}), we have 
\begin{equation} 
f(\rho) \approx \frac{2C_2}{3 C_1} \rho - \frac{C_2^2}{3 C_1} \rho^2 \,.
\label{eq:3.8}
\end{equation}
Since the value of $N$ given by Eq.~(\ref{eq:3.3}) has to be 
positive, we find $C_2 <0$. 
In addition, the number of $e$-folds $N$ during inflation has to 
be much larger than unity such as $N = 60$, and hence, 
from Eq.~(\ref{eq:3.3}) and the condition $\left| C_2\rho \right| \gg 1$, 
we acquire $\left(-C_2 \right)/C_1 \approx 1/N \ll 1$. {}From Eqs.~(\ref{eq:3.7}) and (\ref{eq:3.8}), we get 
\begin{equation} 
w = \frac{P}{\rho} \approx -1 - \frac{2}{3} \left( -\frac{C_2}{C_1} \right) 
+ \frac{1}{3} \left( -\frac{C_2}{C_1} \right) \left(-C_2 \rho \right) 
\approx -1 + \frac{1}{3N} \left(-2-C_2 \rho \right)\,, 
\label{eq:FR16-5-IIIC1-1}
\end{equation}
where in deriving the second approximate equality, we have used $\left(-C_2 \right)/C_1 \approx 1/N$. 
For example, if $\left| C_2\rho \right| = \mathcal{O}(10)$ and 
$\left(-C_2 \right)/C_1 \approx 1/N$, where, e.g., $N \gtrsim 60$, from 
Eq.~(\ref{eq:FR16-5-IIIC1-1}), we can obtain $w \approx -1$. 
This implies that the slow-roll inflation, namely, the de Sitter inflation, 
can occur, and hence the scale factor can be represented as 
\begin{equation} 
a(t) = a_\mathrm{i} \exp \left[ H_\mathrm{inf} (t-t_\mathrm{i}) \right] \,. 
\label{eq:FR16-5-IIIC1-2}
\end{equation}
Moreover, if $\left(-C_2 \right)/C_1 < 1/N$, from Eq.~(\ref{eq:FR16-5-IIIB-01}), it is seen that for $N \gtrsim 73$, the tensor-to-scalar ratio $r$ of the density perturbations can meet $r < 0.11$, 
which is consistent with the Planck results. 

Through the comparison between this expression and Eq.~(\ref{eq:2.10}), 
we see that these expressions become equivalent, i.e., 
the linear combination of $\rho$ and $\rho^2$. 
In this case, there are two combinations of the model parameters, 
which will be called as Model (a) and Model (b) as follows 
\begin{equation} 
\mathrm{Model \,\,\, (a)}: \quad \quad 
A = \frac{2C_2}{3 C_1}\,, 
\quad 
\bar{\zeta} = - \frac{3C_2^2}{C_1 \kappa^4}\,, 
\quad 
\beta =1\,, 
\quad 
\gamma=4 \,,
\label{eq:3.9}
\end{equation}
and 
\begin{equation}
\mathrm{Model \,\,\, (b)}: \quad \quad  
A = -\frac{C_2^2}{3 C_1}\,, 
\quad 
\bar{\zeta} = \frac{2C_2}{C_1 \kappa^2}\,, 
\quad 
\beta =2\,, 
\quad 
\gamma=2 \,. 
\label{eq:3.10}
\end{equation}
In (\ref{eq:3.9}) and (\ref{eq:3.10}), 
when the second relations have been derived by using the forth relations. 
As a result, the EoS of a fluid can explicitly be reconstructed. 

\subsubsection{Case (ii): $\left| C_2\rho \right| \ll 1$}

On the other hand, if 
$\left| C_2\rho \right| \ll 1$, it follows from Eq.~(\ref{eq:3.6}) that 
\begin{equation} 
f(\rho) \approx 
-\frac{1}{3C_1} + \frac{2C_2}{3 C_1} \rho \,.
\label{eq:3.11}
\end{equation}
With Eq.~(\ref{eq:3.3}) and the condition $\left| C_2\rho \right| \ll 1$, 
we have $C_1 \rho \approx N \gg 1$, and 
eventually we also find $\left|C_2\right|/C_1 \ll 1$. {}From Eqs.~(\ref{eq:3.7}) and (\ref{eq:3.8}), we acquire 
\begin{equation} 
w = \frac{P}{\rho} \approx -1 - \frac{1}{3} \frac{1}{C_1 \rho} 
+ \frac{2}{3} \left( \frac{C_2}{C_1} \right) 
\approx -1+ \frac{1}{3} \left( -\frac{1}{N} + 2 \frac{C_2}{C_1} \right)\,. 
\label{eq:FR16-5-IIIC2-1}
\end{equation}
Here, the second approximate equality follows from $C_1 \rho \approx N$. 
Accordingly, from Eq.~(\ref{eq:FR16-5-IIIC2-1}) 
with $1/N \ll 1$ and $\left|C_2\right|/C_1 \ll 1$, 
we see that $w \approx -1$ can be met.  
Thus, the slow-roll (de Sitter) inflation can happen, and 
the scale factor can be expressed by Eq.~(\ref{eq:FR16-5-IIIC1-2}). 
In addition, for $C_2 >0$ and $C_2/C_1 \lesssim 1/N$, 
by using Eq.~(\ref{eq:FR16-5-IIIB-01}), it is found that even for 
$N \gtrsim 60$, the tensor-to-scalar ratio $r$ of the density perturbations 
becomes $r < 0.11$, which is consistent with the Planck results. 
On the other hand, for $C_2 <0$ and $\left|C_2 \right|/C_1 < 1/N$, 
it follows from Eq.~(\ref{eq:FR16-5-IIIB-01}) 
that for $N \gtrsim 73$, we can get $r < 0.11$, 
similarly to that in Case (i) described above. 

The comparison of this expression with Eq.~(\ref{eq:2.10}) leads to 
the following combinations of the model parameters, 
which will be named Model (c) and Model (d) as follows 
\begin{equation} 
\mathrm{Model \,\,\, (c)}: \quad \quad  
A = -\frac{1}{3 C_1}\,, 
\quad 
\bar{\zeta} = \frac{2C_2}{C_1 \kappa^2}\,, 
\quad 
\beta =0\,, 
\quad 
\gamma=2 \,,
\label{eq:3.12}
\end{equation}
and 
\begin{equation} 
\mathrm{Model \,\,\, (d)}: \quad \quad  
A = \frac{2C_2}{3 C_1}\,, 
\quad 
\bar{\zeta} = -\frac{1}{3 C_1}\,, 
\quad 
\beta =1\,, 
\quad 
\gamma=0 \,. 
\label{eq:3.13}
\end{equation}
In Eqs.~(\ref{eq:3.12}) and (\ref{eq:3.13}), with the forth relations, 
the second ones have been derived. 
In Table~\ref{table-1}, the fluid models with the EoS in Eq.~(\ref{eq:3.7}) satisfying the relation (\ref{eq:FR16-4-IIIB-1}) are summarized.

We remark that if $C_2 >0$, the inflaton potential can correspond to 
the one in the Starobinsky inflation. {}From the investigations 
in the scalar field theories, 
we have $C_2 = \left(2/3\right) C_1$~\cite{Chiba:2015zpa}. 
In this case, for the models in Eqs.~(\ref{eq:3.12}) and (\ref{eq:3.13}), 
we obtain $\bar{\zeta} = 4/\left(3 \kappa^2 \right)$ and 
$A = 4/9$, respectively.

\begin{table}[t]
\caption{Fluid models with the EoS in Eq.~(\ref{eq:3.7}) realizing 
the relation (\ref{eq:FR16-4-IIIB-1}). Here, $C_1 >0$. 
In Case (i), $\left| C_2\rho \right| \gg 1$ and $C_2 <0$, 
whereas for Case (ii), $\left| C_2\rho \right| \ll 1$ and $C_2$ can take 
both the positive and negative values. 
}
\begin{center}
\tabcolsep = 2mm 
\begin{tabular}
{cccccc}
\hline
\hline 
Case 
& Model 
& $A$  
& $\bar{\zeta}$  
& $\beta$ 
& $\gamma$ 
\rule{0mm}{3.5mm}
\\[0mm]
\hline
(i)
&(a) 
& $2C_2/\left(3 C_1\right)$
& $-3C_2^2/\left(C_1 \kappa^4 \right)$
& $1$ 
& $4$  
\\[0mm]
(i)
&(b)  
& $-C_2^2/\left(3 C_1\right)$
& $2C_2/\left(C_1 \kappa^2 \right)$
& $2$ 
& $2$  
\\[0mm]
(ii)
&(c)  
& $-1/\left(3 C_1\right)$
& $2C_2/\left(C_1 \kappa^2 \right)$
& $0$ 
& $2$ 
\\[0mm]
(ii)
&(d)  
& $2C_2/\left(3 C_1\right)$
& $-1/\left(3 C_1\right)$
& $1$ 
& $0$  
\\[1mm]
\hline
\hline
\end{tabular}
\end{center}
\label{table-1}
\end{table}

\section{Graceful exit from inflation} 

In this section, 
we examine whether the graceful exit from inflation 
can occur in a fluid model. 
We analyze the instability of the de Sitter solution 
($H = H_\mathrm{inf} \, (>0) = \mathrm{constant}$) 
during inflation by taking the perturbations of 
the Hubble parameter as follows~\cite{Bamba:2015uxa} 
\begin{equation}
H = H_\mathrm{inf} + H_\mathrm{inf} \delta(t) \,. 
\label{eq:4.1} 
\end{equation}
Here, $\left| \delta(t) \right| \ll 1$, and 
hence $H_\mathrm{inf} \delta(t)$ denotes 
the perturbations from the de Sitter solution $H_\mathrm{inf}$. 

We rewrite Eq.~(\ref{eq:II-6}) as the following second differential equation 
with respect to the cosmic time $t$: 
\begin{equation}
\ddot{H} -\frac{\kappa^4}{2}\left[\beta A^2 \left( \frac{3}{\kappa^2} \right)^{2\beta} H^{4\beta-1} + \left(\beta + \frac{\gamma}{2} \right) A \bar{\zeta} \left( \frac{3}{\kappa^2} \right)^{\beta} H^{2\beta + \gamma - 1} + 
\frac{\gamma}{2} \bar{\zeta}^2 H^{2\gamma - 1} \right] = 0 \,.
\label{eq:4.2} 
\end{equation}
We define the form of $\delta(t)$ as 
\begin{equation} 
\delta(t) \equiv \e^{\lambda t} \,, 
\label{eq:4.3} 
\end{equation}
where $\lambda$ is a constant, so that we can investigate the 
instability of the de Sitter solution. 
If there is a positive solution of $\lambda$, 
the de Sitter solution can be unstable. 
Therefore, the universe can exit from inflation, and 
the reheating stage can follow, because 
the absolute value of $\delta(t)$ with $\lambda > 0$ 
becomes larger as the cosmic time grows at the inflationary stage. 

We substitute Eq.~(\ref{eq:4.1}) with Eq.~(\ref{eq:4.3}) 
into Eq.~(\ref{eq:4.2}) and take the first order of $\delta (t)$. 
Accordingly, we get 
\begin{eqnarray} 
&&
\lambda^2 - \frac{1}{2} \frac{\kappa^4}{H_\mathrm{inf}^2} \mathcal{Q} = 0 \,, 
\label{eq:4.4} \\
&&
\mathcal{Q} \equiv 
\beta \left(4\beta - 1\right) A^2 \left( \frac{3}{\kappa^2} \right)^{2\beta} 
H_\mathrm{inf}^{4\beta} + \left(\beta + \frac{\gamma}{2} \right) 
\left(2\beta + \gamma - 1\right) A \bar{\zeta} 
\left( \frac{3}{\kappa^2} \right)^{\beta} H_\mathrm{inf}^{2\beta + \gamma} + 
\frac{\gamma}{2} \left(2\gamma - 1\right) 
\bar{\zeta}^2 H_\mathrm{inf}^{2\gamma} \,.
\label{eq:4.5}
\end{eqnarray}
We see that the solutions of Eq.~(\ref{eq:4.4}) are given by 
\begin{equation} 
\lambda = \lambda_\pm \equiv \pm \frac{1}{\sqrt{2}} \frac{\kappa^2}{H_\mathrm{inf}} 
\sqrt{\mathcal{Q}}\,. 
\label{eq:4.6} 
\end{equation}
If $\mathcal{Q} > 0$, we can acquire the positive solution of $\lambda = \lambda_+ > 0$. 
As a result, the exit from inflation can gracefully occur. 

Concretely, in the fluid models reconstructed above and summarized in Table~\ref{table-1}, we check whether the graceful exit from inflation can be realized or not, namely, whether $\mathcal{Q}$ can take a positive value or not. 
If the universe cannot successfully exit from inflation, 
inflation does not ends, and therefore such a scenario corresponds to the so-called eternal inflation. 
By substituting the values of $A$, $\bar{\zeta}$, $\beta$, and $\gamma$ 
in Models (a), (b), (c), and (d), given by Eqs.~(\ref{eq:3.9}), (\ref{eq:3.10}), (\ref{eq:3.12}), and (\ref{eq:3.13}), respectively, 
into Eq.~(\ref{eq:4.6}), we obtain the expressions of $\mathcal{Q}$ 
in each models. 
To evaluate the values of $\mathcal{Q}$, 
we take into account the following facts. 
For all of the models, $C_1 >0$. 
On the other hand, in Models (a) and (b), $C_2 <0$, 
while in Models (c) and (d), 
$C_2$ can become both the positive and negative values. 
For Models (a) and (b) in Case (i), we find
\begin{eqnarray} 
&&
\mathrm{Model \,\,\, (a)}: \quad \quad  
\mathcal{Q} = 2 \left(\frac{C_2}{C_1}\right)^2 
\left(\frac{H_\mathrm{inf}}{\kappa}\right)^4 
\left[ 6 - 45 C_2 \left(\frac{H_\mathrm{inf}}{\kappa}\right)^2 
+ 63 C_2^2 \left(\frac{H_\mathrm{inf}}{\kappa}\right)^4 \right] >0 \,,
\label{eq:4.7} \\ 
&&
\mathrm{Model \,\,\, (b)}: \quad \quad  
\mathcal{Q} = 6 \left(\frac{C_2}{C_1}\right)^2 
\left(\frac{H_\mathrm{inf}}{\kappa}\right)^4 
\left[ 2 - 15 C_2 \left(\frac{H_\mathrm{inf}}{\kappa}\right)^2 
+ 21 C_2^2 \left(\frac{H_\mathrm{inf}}{\kappa}\right)^4 \right] >0 \,.
\label{eq:4.8} 
\end{eqnarray}
Accordingly, we always have $\mathcal{Q} >0$. 
While, for Models (c) and (d) in Case (ii), we acquire
\begin{eqnarray} 
&&
\mathrm{Model \,\,\, (c)}: \quad \quad  
\mathcal{Q} = \left(\frac{C_2}{C_1}\right)^2 
\left(\frac{H_\mathrm{inf}}{\kappa}\right)^2
\left[ - \frac{1}{3C_2} + 12 \left(\frac{H_\mathrm{inf}}{\kappa}\right)^2 
\right] \,,
\label{eq:4.9} \\ 
&&
\mathrm{Model \,\,\, (d)}: \quad \quad  
\mathcal{Q} = 2\left(\frac{C_2}{C_1}\right)^2 
\left(\frac{H_\mathrm{inf}}{\kappa}\right)^2
\left[ 6 \left(\frac{H_\mathrm{inf}}{\kappa}\right)^2 
- \frac{1}{3C_2} \right] \,.
\label{eq:4.10} 
\end{eqnarray}
{}From these relations, we find that if $C_2 <0$, we get $\mathcal{Q} >0$, 
whereas, in the case that $C_2 >0$, 
if the following conditions are satisfied
\begin{eqnarray} 
C_2 \Eqn{>} \frac{1}{36} \left(\frac{\kappa}{H_\mathrm{inf}}\right)^2 
\quad \quad 
\mathrm{for} \,\,\, 
\mathrm{Model \,\,\, (c)}\,, 
\label{eq:4.11} \\
C_2 \Eqn{>} \frac{1}{18} \left(\frac{\kappa}{H_\mathrm{inf}}\right)^2 
\quad \quad 
\mathrm{for}  \,\,\, 
\mathrm{Model \,\,\, (d)}\,, 
\label{eq:4.12} 
\end{eqnarray}
we can obtain $\mathcal{Q} >0$. 
Thus, for the reconstructed models of a fluid in the previous section, it is possible for the universe to gracefully exit from inflation. 

In table~\ref{table-2}, we present the summary of the reconstructed fluid models. We show the EoS of these models in the form of Eq.~(\ref{eq:2.8}) so that 
a term inspired by bulk viscosity can clearly be seen. 
In these models, the three observables of inflationary cosmology 
can be compatible with the Planck results. 
First, 
the spectral index $n_\mathrm{s}$ of the curvature perturbations 
is expressed as $n_\mathrm{s} - 1 = -2/N$ in Eq.~(\ref{eq:FR16-4-IIIB-1}), 
which can lead to $0.967$ for $N=60$. 
Second, 
The tensor-to-scalar ratio $r$ of the density perturbations 
can satisfy the upper limit of $r < 0.11$. 
In Models (a) and (b) [Case (i)] and 
Models (c) and (d) [Case (ii)] with $C_2 <0$, 
if $N \gtrsim 73$, we can obtain $r < 0.11$. 
While, in Models (c) and (d) [Case (ii)] with $C_2 >0$, 
when $N \gtrsim 60$, we can find $r < 0.11$. 
Third, the running $\alpha_\mathrm{s}$ of the spectral index is 
given by $\alpha_\mathrm{s} = -2/N^2$ in Eq.~(\ref{eq:FR16-7-IIIB-001}). 
From this expression, we have $\alpha_\mathrm{s} = -5.56 \times 10^{-4}$. 
These values of $n_\mathrm{s}$, $r$, and $\alpha_\mathrm{s}$ 
are consistent with the Planck results. 
Moreover, the universe can gracefully exit from inflation. 
We describe the conditions for the graceful exit from inflation. 

\begin{table}[t]
\caption{The EoS for the fluid models reconstructed 
in Sec.~III and the conditions that in these models, 
the graceful exit from inflation can be realized. 
In these models, 
$n_\mathrm{s} - 1 = -2/N = 0.967$ for $N=60$, 
$r < 0.11$ for $N \gtrsim 73$ in Models (a), (b) and 
Models (c) and (d) with $C_2 <0$, or $N \gtrsim 60$ 
in Models (c) and (d) with $C_2 >0$, 
and $\alpha_\mathrm{s} = -2/N^2 = -5.56 \times 10^{-4}$ 
for $N=60$ can be realized. These values can explain 
the Planck data. 
Legend is the same as Table~\ref{table-1}.}
\begin{center}
\tabcolsep = 2mm 
\begin{tabular}
{cccc}
\hline
\hline 
Case 
& Model 
& EoS  
& Conditions for the graceful exit from inflation 
\\[0mm]
\hline
(i)
&(a) 
& $P=-\rho+ \left[2C_2/\left(3 C_1\right) \right]\rho 
- \left[3C_2^2/\left(C_1 \kappa^4 \right)\right] H^4$
& No condition
\rule{0mm}{3.5mm}
\\[0mm]
(i)
&(b)  
& $P=-\rho -\left[ C_2^2/\left(3 C_1\right) \right]\rho^2 
+ \left[2C_2/\left(C_1 \kappa^2 \right) \right] H^2$
& No condition
\\[0mm]
(ii)
&(c)  
& $P=-\rho - \left[1/\left(3 C_1\right) \right] 
+ \left[2C_2/\left(C_1 \kappa^2 \right) \right] H^2$
& $C_2 <0$ or $C_2 > \left(1/36\right) 
\left(\kappa/H_\mathrm{inf}\right)^2$
\\[0mm]
(ii)
&(d)  
& $P=-\rho + \left[2C_2/\left(3 C_1\right) \right] \rho
- \left[1/\left(3 C_1\right) \right]$
& $C_2 <0$ or $C_2 > \left(1/18\right) 
\left(\kappa/H_\mathrm{inf}\right)^2$
\\[1mm]
\hline
\hline
\end{tabular}
\end{center}
\label{table-2}
\end{table}

\section{Singular inflation in a fluid model} 

In this section, we study the singular inflation~\cite{Nojiri:2015wsa} 
in a fluid model. 
In this inflationary scenario, 
the idea of finite-time future singularities in the context of 
the dark energy problem is applied to inflation in the early universe. 

The finite-time future singularities are classified into four types~\cite{Nojiri:2005sx}. Their features in modified gravity theories 
have also been analyzed in detail~\cite{Bamba:2008ut} 
(for a detailed review on the finite-time future singularities, see~\cite{Bamba:2012cp}). 
Among them, the formulation of the Type IV singularity can be used in the singular inflation, because there is no divergence in terms of the scale factor, the energy density and pressure of the universe. 

In the Type IV singularity, for $t\to t_{\mathrm{s}}$, where $t_{\mathrm{s}}$ is the time when the singularity appears, 
the scale factor $a$, the effective (i.e., total) energy density $\rho_{\mathrm{eff}}$ and pressure $P_{\mathrm{eff}}$ of the universe become finite 
as $a \to a_{\mathrm{s}}$, 
$\rho_{\mathrm{eff}} \to 0$ 
and $\left| P_{\mathrm{eff}} \right| \to 0$. 
Here, $a_{\mathrm{s}}$ is the value of $a$ at $t=t_{\mathrm{s}}$. 
The case that $\rho_{\mathrm{eff}}$ and/or $\left| P_{\mathrm{eff}} \right|$
become non-zero finite values at $t = t_{\mathrm{s}}$~\cite{SS-B} is also included in the Type IV singularity. 
However, the higher derivatives of $H$ diverge. 

In the following, we explore the inflationary stage in which 
there is only a component of a fluid. 
Therefore, for simplicity, we describe $\rho_{\mathrm{eff}}$ and $P_{\mathrm{eff}}$ by $\rho$ and $P$, respectively. 
We consider the case that the Hubble parameter and scale factor during inflation are represented as 
\begin{eqnarray}
H \Eqn{=} H_\mathrm{inf} + \bar{H}\left(t_{\mathrm{s}} -t \right)^{q} \,, 
\quad q>1 \,,
\label{eq:5.1}\\
a \Eqn{=} \bar{a} \exp \left[ H_\mathrm{inf}t -\frac{\bar{H}}{q+1} \left(t_{\mathrm{s}} -t \right)^{q+1} \right] \,, 
\label{eq:5.2} 
\end{eqnarray}
where $\bar{H}$, $q$, and $\bar{a}$ are constants, and the mass dimension of 
$\bar{H}$ is $q+1$. 

In the flat FLRW universe, 
from the gravitational field equations, the energy density and pressure of 
the universe are given by 
\begin{equation} 
\rho = \frac{3H^2}{\kappa^2}\,, 
\quad 
P = -\frac{2\dot{H}+3H^2}{\kappa^2} \,.
\label{eq:5.3} 
\end{equation}
It is seen from Eq.~(\ref{eq:5.2}) and 
the expressions in (\ref{eq:5.3}) with Eq.~(\ref{eq:5.1}) 
that in the limit $t\to t_{\mathrm{s}}$, all of $a$, $\rho$, and $P$ 
asymptotically approach finite values, 
while the higher derivatives of $H$ diverge. 
Thus, the Type IV singularity appears at $t = t_{\mathrm{s}}$. 
By using the expressions of $\rho$ and $P$ in (\ref{eq:5.2}) 
with Eq.~(\ref{eq:5.1}), we find the following EoS for a fluid 
\begin{equation} 
P = -\rho + f(\rho) \,, 
\quad 
f(\rho) = 
\frac{2q \bar{H}^{1/q}}{\kappa^2} \left( \kappa \sqrt{\frac{\rho}{3}} - H_\mathrm{inf} \right)^{\left(q-1\right)/q} \,.
\label{eq:5.4} 
\end{equation}
Here, $f(\rho)$ can be described as the series of power of $\rho$. 
In fact, if $H_\mathrm{inf}/\sqrt{\kappa^2 \rho/3} = H_\mathrm{inf}/H \ll 1$, where the equality comes from the first equation in (\ref{eq:5.3}), 
we find 
\begin{equation} 
f(\rho) \approx \frac{2}{3^{\left(q-1\right)/\left(2q\right)}} 
\frac{\bar{H}^{1/q}}{\kappa^{\left(q+1\right)/q}} 
\left[ \rho^{\left(q-1\right)/\left(2q\right)} 
- \frac{\sqrt{3}\left(q-1\right)}{q} 
\frac{H_\mathrm{inf}}{\kappa} \rho^{-1/\left(2q\right)} \right] \,,
\label{eq:5.5} 
\end{equation}
where we have taken the first order of the quantity 
$\left(H_\mathrm{inf}/\sqrt{\left(\kappa^2 \rho\right)/3}\right)$. 
It follows from Eq.~(\ref{eq:5.5}) that $f(\rho)$ is 
represented as a linear combination of two powers of $\rho$, 
similarly to that in Eq.~(\ref{eq:2.10}) or Eq.~(\ref{eq:3.8}). 
Hence, this model can be regarded as a kind of 
the fluid models reconstructed in Sec.~III. 

Furthermore, Eq.~(\ref{eq:5.5}) divided by $\rho$ reads 
\begin{equation} 
\frac{f(\rho)}{\rho} \approx 
\frac{2q}{3} \bar{H}^{1/q} \left(\frac{\kappa^2 \rho}{3}\right)^{-\left(q+1\right)/\left(2q\right)}  
\left[ 1 - \frac{\left(q-1\right)}{q} 
\frac{H_\mathrm{inf}}{\sqrt{\kappa^2 \rho/3}}\right] 
= \frac{2q}{3} \left(\frac{\bar{H}}{H^{q+1}}\right)^{1/q} 
\left[ 1 - \frac{\left(q-1\right)}{q} \frac{H_\mathrm{inf}}{H} \right]\,. 
\label{eq:5.6} 
\end{equation}
Here, in deriving the last equality, we have used the first equation 
in (\ref{eq:5.3}). It is seen from Eq.~(\ref{eq:5.6}) that for $\bar{H}/H^{q+1} \ll 1$, we have $f(\rho)/\rho \ll 1$. In this case, 
the observables of the inflationary models, i.e., 
$n_\mathrm{s}$, $r$, and $\alpha_\mathrm{s}$, can approximately be 
represented by Eq.~(\ref{eq:2.6}), and 
the values of $n_\mathrm{s}$, $r$, and $\alpha_\mathrm{s}$ can be 
compatible with the Planck analysis, as stated in Sec.~II C. 

We explain the existence of limit of $\bar{\zeta} = 0$ in Eq.~(\ref{eq:2.9}), 
in which the term $\zeta(H)$ in Eq.~(\ref{eq:2.9}) will not exist 
and therefore the EoS for a fluid in Eq.~(\ref{eq:2.8}) reads 
$P=-\rho + A\rho^{\beta}$. In such a limit, from Eq.~(\ref{eq:2.5}), 
we have $f(\rho) = A\rho^{\beta}$, namely, the term $f(\rho)$ consists of 
the single power of $\rho$. 
On the other hand, in Eqs.~(\ref{eq:5.5}) and (\ref{eq:5.6}), 
the form of $f(\rho)$ is a linear combination of 
two kinds of power of $\rho$. 
The form of $f(\rho)$ can approximately be given in Eqs.~(\ref{eq:5.5}) and (\ref{eq:5.6}) only if the singular inflation occurs 
and the Hubble parameter and the scale factor are 
expressed as Eqs.~(\ref{eq:5.1}) and (\ref{eq:5.2}), respectively, 
This means that for a fluid without the term $\zeta(H)$ in Eq.~(\ref{eq:2.9}), 
the singular inflation cannot be realized. 
Thus, the existence of the term $\zeta(H)$ can influence 
the dynamics of the universe filled with a fluid in the early universe. 

As a consequence, it is considered that 
the singular inflation can be realized in the fluid models 
in which the spectral index of the curvature perturbations 
can explain the recent Planck results.

\section{Equivalence between a fluid description of inflation and the description of inflation in terms of scalar field theories}

In this section, we demonstrate that a fluid description of inflation can be equivalent to the description of inflation in terms of scalar field theories (for further related investigations, see Ref.~\cite{Bamba:2012cp}). 
The action of scalar field theories is expressed as 
\begin{equation} 
S = \int d^4 x \sqrt{-g} \left( \frac{R}{2\kappa^2} - 
\frac{1}{2} \omega (\varphi) {\partial}_{\mu} \varphi {\partial}^{\mu} \varphi 
- V(\varphi) \right)\,. 
\label{eq:6.1}
\end{equation}
Here, 
$\omega (\varphi)$ is a coefficient function of kinetic term 
of the scalar field $\varphi$ and $V(\varphi)$ is the potential of $\varphi$. 
Starting from a fluid description, 
we construct a scalar field theory with the same 
EoS as that of a fluid. 
By this process, we obtain the expressions of 
$\omega (\varphi)$ and $V(\varphi)$ 
of the corresponding scalar field theory to a fluid description. 
Consequently, we can represent a fluid description 
as the description of a scalar field theory. 

It is known that in the FLRW background, $\omega (\varphi)$ and $V(\varphi)$ can be described as~\cite{Capozziello:2005mj}
\begin{eqnarray}
\omega (\varphi) \Eqn{=} -\frac{2}{\kappa^2}
\frac{dJ(\varphi)}{d \varphi}\,,
\label{eq:6.2} \\
V(\varphi) \Eqn{=} \frac{1}{\kappa^2}
\left[ 3\left(J (\varphi)\right)^2 + \frac{dJ(\varphi)}{d \varphi} \right]\,, 
\label{eq:6.3}
\end{eqnarray} 
with $J(\varphi)$ an arbitrary function of $\varphi$. 
Here, we can take $\varphi = t$ and $H= J(t)$ because 
$\varphi$ can be treated as an auxiliary scalar quantity. 
On the other hand, the energy density $\rho$ and pressure $P$ of the 
scalar field $\varphi$ read 
\begin{eqnarray}
\rho \Eqn{=} \frac{1}{2} \omega (\varphi) \dot{\varphi}^2
+V(\varphi)\,,
\label{eq:6.4} \\
P \Eqn{=} \frac{1}{2} \omega (\varphi) \dot{\varphi}^2
-V(\varphi)\,. 
\label{eq:6.5}
\end{eqnarray}
With these equations, we find that $\omega (\varphi)$ and $V(\varphi)$ 
are given by 
\begin{eqnarray}
\omega \dot{\varphi}^2 \Eqn{=} \rho + P = f(\rho)\,,
\label{eq:6.6} \\ 
V \Eqn{=} \frac{1}{2} \left( \rho - P \right) = \rho - \frac{f(\rho)}{2}\,.
\label{eq:6.7} 
\end{eqnarray}
In deriving the second equalities in Eqs.~(\ref{eq:6.6}) and (\ref{eq:6.7}), 
we have used the EoS of $P=-\rho + f(\rho)$ 
in a fluid description in Eq.~(\ref{eq:2.5}). {}From the Friedmann equation (\ref{eq:FR16-3-IIB1}), we find $\rho = 3H^2/\kappa^2$. Therefore, when we have $H (= I(t))$, we can express $\rho = \rho (t (\varphi)) = \rho (\varphi)$ as a function of $t (= \varphi)$. Eventually, from Eqs.~(\ref{eq:6.6}) and (\ref{eq:6.7}) with $\rho = \rho (\varphi)$, we can acquire the expressions of 
$\omega = \omega (\varphi)$ and $V = V (\varphi)$. 
By using these processes, we can obtain the description of 
a scalar field theory corresponding to an original fluid description. 

Moreover, we consider the opposite approach from the description of a scalar field to a fluid description. We first have a scalar field action with 
$\omega (\varphi)$ and $V (\varphi)$ in Eq.~(\ref{eq:6.1}). 
It follows from Eqs.~(\ref{eq:6.4}) and (\ref{eq:6.5}) with $\phi = t$ and $H = J(t)$ that the EoS of the universe 
$w \equiv P/\rho = -1 + f(\rho)/\rho$ in Eq.~(\ref{eq:6.1}) with Eq.~(\ref{eq:2.5}). 
Plugging this relation with Eq.~(\ref{eq:6.4}) and 
comparing the relation obtained with $w = -1 + f(\rho)/\rho$, 
we get $f(\rho)$ in a fluid description. 
Thus, it can be considered that 
both approaches shown above suggests the equivalence between 
a fluid description and the description in terms of 
scalar field theories. 

As a fluid description, we the following case that 
$f(\rho)$ is given by 
\begin{equation} 
f(\rho) = \bar{f}_1 + \bar{f}_2 \left(\frac{\rho}{\rho_*}\right)^u \,, 
\label{eq:6.8}
\end{equation}
where $\bar{f}_1$, $\bar{f}_2$, and $u$ are constants, 
and $\rho_*$ is a fiducial value of $\rho$
For $f(\rho)$ in Eq.~(\ref{eq:6.8}) with $u = 1$, 
we have the following EoS 
\begin{equation} 
P = -\rho + \left(\frac{\bar{f}_2}{\rho_*}\right) \rho + \bar{f}_1 \,, 
\label{eq:6.9}
\end{equation}
By comparing this form with Model (d) in Table~\ref{table-2}, 
it is seen that if $\bar{f}_2/\rho_* = 2C_2/\left(3 C_1\right)$ 
and $\bar{f}_1 = -\left[1/\left(3 C_1\right) \right]$, 
the form of the EoS in Eq.~(\ref{eq:6.9}) is 
equal to Model (d). 
This means that in a fluid description with the form of 
the EoS in Eq.~(\ref{eq:6.8}), 
the spectral index of the curvature perturbations, 
the tensor-to-scalar ratio of the density perturbations, 
and the running of the spectral index 
can be consistent with the Planck results. 

{}From the substitution of Eq.~(\ref{eq:6.8}) 
into the relation $w = -1 + f(\rho)/\rho$, Eqs.~(\ref{eq:6.6}) and (\ref{eq:6.7}), we acquire 
\begin{eqnarray}
w \Eqn{=} -1 + \frac{\bar{f}_1}{\rho} +\frac{\bar{f}_2}{\rho_*} 
\left(\frac{\rho}{\rho_*}\right)^{u-1}\,, 
\label{eq:6.10} \\
\omega \dot{\varphi}^2 \Eqn{=}  
\bar{f}_1 + \bar{f}_2 \left(\frac{\rho}{\rho_*}\right)^u \,,
\label{eq:6.11} \\ 
V \Eqn{=} \rho - \frac{1}{2} \left[ 
\bar{f}_1 + \bar{f}_2 \left(\frac{\rho}{\rho_*}\right)^u \right]\,.
\label{eq:6.12} 
\end{eqnarray}

In addition, for instance, we consider the case that 
the Hubble parameter and the scale factor during inflation are expressed by 
\begin{eqnarray} 
H \Eqn{=} \frac{\bar{h}}{t}\,,
\label{eq:6.13}\\
a \Eqn{=} \tilde{a} t^{\bar{h}}\,,
\label{eq:6.14}
\end{eqnarray}
where $\bar{h} \gg 1$ and $\tilde{a} (\neq 0)$ are constants. 
Such a case of $\bar{h} \gg 1$ corresponds to 
the quasi-de Sitter inflation (i.e., the slow-roll inflation). 
In this case, it follows from Eqs.~(\ref{eq:6.2}) and (\ref{eq:6.3}) with 
$\varphi = t$ and $H= J(t)$ that 
\begin{eqnarray}
\omega (\varphi) \Eqn{=}  
\frac{2\bar{h}}{\kappa^2} \frac{1}{\varphi^2}\,,
\label{eq:6.15} \\ 
V (\varphi) \Eqn{=} \frac{\bar{h}\left(3\bar{h}-1\right)}{\kappa^2} \frac{1}{\varphi^2}\,.
\label{eq:6.16} 
\end{eqnarray}
As described above, 
when the Hubble parameter $H$ can be represented as a function of $t$, {}from 
Eqs.~(\ref{eq:6.2}) and (\ref{eq:6.3}) with $\varphi = t$ and $H= J(t)$, 
the expressions of $\omega = \omega (\varphi)$ and 
$V = V (\varphi)$ can be derived explicitly.

\section{Conclusions}

In the present paper, we have investigated 
the description of the inflationary universe in the framework of 
a fluid model in which the EoS for a fluid includes bulk viscosity. 
It has been found that in a fluid description, 
the three observables of inflationary models, 
namely, the spectral index $n_\mathrm{s}$ of the curvature perturbations, 
the tensor-to-scalar ratio $r$ of the density perturbations, 
and the running $\alpha_\mathrm{s}$ of the spectral index, 
can be consistent with the recent Planck results. 

Furthermore, 
we have explicitly reconstructed the EoS of a fluid model 
from the spectral index $n_\mathrm{s}$ of the curvature perturbations. 
Particularly, we have used the expression of $n_\mathrm{s}$ as a function of the number of $e$-folds $N$ 
in the inflationary models, where the value of $n_\mathrm{s}$ 
can explain the Planck data, including the Starobinsky inflation. 
It has been shown that 
for the fluid models reconstructed from the spectral index, 
indeed, the slow-roll (de Sitter) inflation can occur. 
It has also been certified that in these fluid models, 
the tensor-to-scalar ratio $r$ of the density perturbations 
can meet the upper limit found by the Planck analysis. 
The running $\alpha_\mathrm{s}$ of the spectral index can be 
compatible with the Planck results. 

In our previous work~\cite{Bamba:2014wda}, 
since we have considered a fluid without 
the term $\zeta(H)$ in Eq.~(\ref{eq:2.9}), 
only for the special case that the EoS for a fluid is 
approximately equal to $-1$ as $w = P/\rho \approx -1$, 
it has been shown that in a fluid model, 
the three observables of inflationary models can be 
consistent with the Planck results. 
On the other hand, in this work, we have introduced 
the additional term $\zeta(H)$ in Eq.~(\ref{eq:2.9}) 
into the EoS for a fluid as in Eq.~(\ref{eq:2.8}). 
As a result, it has been found that also for cases 
in which the value of the EoS $w$ for a fluid is apart from 
$-1$, in such a fluid model, the three observables of inflationary models 
can be compatible with the Planck analysis. 

In addition, we have examined the instability of the de Sitter solution 
at the inflationary stage by analyzing the perturbations of 
the Hubble parameter. 
It has been performed that 
the universe can gracefully exit from inflation in the reconstructed 
models of a fluid. 
We have also derived the conditions for the graceful exit from 
inflation to be realized in the reconstructed fluid models. 

Moreover, we have explored the singular inflation in a fluid model 
by using the formulations to describe the Type IV singularity, 
which is one of the four types of the finite-time future singularity. 
It has been demonstrated that the singular inflation can be realized 
in the fluid models where the spectral index of the curvature perturbations can be compatible with the Planck data. 

It has also been studied that a fluid description of inflation can be equivalent to the description of inflation in terms of scalar field theories. 

Consequently, not only the representation of inflation in scalar field 
theories but also a fluid description of the inflationary universe 
can explain the observational results acquired by the Planck satellite. 

The present method of the reconstruction may equally be applied in the case 
that the universe is filled with several coupled fluids. 
This description may also be applied to the cosmological 
evolution from modified gravity consistent with a fluid description 
at the background evolution level.

\section*{Acknowledgments}

This work was partially supported by the JSPS Grant-in-Aid for 
Young Scientists (B) \# 25800136 and 
the research-funds provided by Fukushima University (K.B.), and 
MINECO (Spain) project FIS2013-44881 (S.D.O.).

\appendix

\section{Slow-roll parameters in a fluid description}

In this Appendix, we present 
the slow-roll parameters in a fluid description. 
The explicit expressions are given by 
\begin{eqnarray}
\epsilon \Eqn{=} 
\frac{3}{2} \rho(N) f(\rho) 
\left( \frac{f'(\rho)-2}
{2\rho(N) - f(\rho)}\right)^2 \, , 
\label{eq:A.1} 
\nonumber \\ 
\eta \Eqn{=} 
\frac{3\rho(N)}{2\rho(N) -f(\rho)} 
\left\{ \frac{f(\rho)}{\rho(N)} + \frac{1}{2} 
\left(f'(\rho)\right)^2 + f'(\rho) 
-\frac{5}{2} \frac{f(\rho)f'(\rho)}{\rho(N)} + 
\left(\frac{f(\rho)}{\rho(N)}\right)^2
\right. \nonumber \\
&& \left. 
{}+\frac{1}{3} \frac{\rho'(N)}{f(\rho)} 
\left[\left(f'(\rho)\right)^2 + f(\rho) f''(\rho) 
-2 \frac{f(\rho)f'(\rho)}{\rho(N)} + \left( \frac{f(\rho)}{\rho(N)} 
\right)^2 \right] 
\right\} \,, 
\label{eq:A.2} \nonumber 
\end{eqnarray}
\begin{eqnarray}
\xi^2 \Eqn{=} 
\frac{f(\rho)\rho(N)\left(f'(\rho) - 2\right)}{2\left(2\rho(N) - f(\rho)\right)^2} \left[ \frac{45}{2} \frac{f(\rho)}{\rho(N)} \left(f'(\rho)-\frac{1}{2} \frac{f(\rho)}{\rho(N)}\right) + 18\left(\frac{f(\rho)}{\rho(N)}\right)^{-1} 
\left(f'(\rho)-\frac{1}{2} \frac{f(\rho)}{\rho(N)}\right)^2 
\right. \nonumber \\
&& \left. 
{}+18\left(\frac{f(\rho)}{\rho(N)}\right)^{-1} 
\left(f'(\rho)-\frac{1}{2} \frac{f(\rho)}{\rho(N)}\right)^3 
-9\left(f'(\rho)-\frac{1}{2} \frac{f(\rho)}{\rho(N)}\right)^2 
-45f'(\rho) + 9\frac{f(\rho)}{\rho(N)} 
+ I \right]
\,,
\label{eq:A.3} \nonumber 
\end{eqnarray}
where 
\begin{eqnarray} 
\hspace{-15mm}
&&
I \equiv 
3 \left(4f'(\rho) -7\frac{f(\rho)}{\rho(N)} +2\right) 
\left\{
-\frac{3}{2}\left(f'(\rho) -\frac{1}{2}\frac{f(\rho)}{\rho(N)}\right) 
\right. \nonumber \\
\hspace{-15mm}
&& \left. 
{}+ \left(\frac{f(\rho)}{\rho(N)}\right)^{-2} 
\frac{\rho'(N)}{\rho(N)} 
\left[ \left(f'(\rho)\right)^2 + f(\rho) f''(\rho) 
-2 \frac{f(\rho)f'(\rho)}{\rho(N)} + \left( \frac{f(\rho)}{\rho(N)} 
\right)^2 \right] \right\} 
\nonumber \\ 
\hspace{-15mm}
&&
{}+2 \left(\frac{f(\rho)}{\rho(N)}\right)^{-2} 
\left\{ 
-\frac{3}{2} \left(\frac{f(\rho)}{\rho(N)}\right) 
\left(\frac{\rho'(N)}{\rho(N)}\right) 
\left[ 
3\left( f'(\rho) \right)^2 
+2f(\rho) f''(\rho) 
-\frac{11}{2}\frac{f(\rho) f'(\rho)}{\rho(N)} 
+\frac{5}{2} \left( \frac{f(\rho)}{\rho(N)} \right)^2
\right]
\right. 
\nonumber \\ 
\hspace{-15mm}
&& \left. 
{}
+ \left(\frac{\rho''(N)}{\rho(N)}\right) 
\left[
\left( f'(\rho) \right)^2 + f(\rho) f''(\rho) 
-2\frac{f(\rho) f'(\rho)}{\rho(N)} 
+ \left( \frac{f(\rho)}{\rho(N)} \right)^2
\right] 
\right. 
\nonumber \\ 
\hspace{-15mm}
&& \left. 
{}
+ \left(\frac{\rho'(N)}{\rho(N)}\right)^2 
\left[ \left(3f'(\rho)f''(\rho) + f(\rho)f'''(\rho)  \right) \rho(N) 
-3\left( f'(\rho) \right)^2 - 3f(\rho) f''(\rho) 
+6\frac{f(\rho) f'(\rho)}{\rho(N)} 
-3\left( \frac{f(\rho)}{\rho(N)} \right)^2 
\right] 
\right\}
\,.
\label{eq:A.4} \nonumber
\end{eqnarray}

\end{document}